\documentclass[notitlepage,a4paper,aps,prd,onecolumn,superscriptaddress,nofootinbib,groupedaddress,longbibliography]{revtex4-1}
\usepackage{amsmath}
\usepackage{amsfonts}
\usepackage{amssymb}
\usepackage{accents}
\usepackage[utf8]{inputenc}
\usepackage[T1]{fontenc}
\usepackage[colorlinks=true]{hyperref}

\newcommand{\dd}{\mathrm{d}}
\newcommand{\lc}[1]{\accentset{\circ}{#1}}
\DeclareMathOperator{\sgn}{sgn}

\begin{document}

\title{General covariant symmetric teleparallel cosmology}

\author{Manuel Hohmann}
\email{manuel.hohmann@ut.ee}
\affiliation{Laboratory of Theoretical Physics, Institute of Physics, University of Tartu, W. Ostwaldi 1, 50411 Tartu, Estonia}

\begin{abstract}
Symmetric teleparallel gravity theories, in which the gravitational interaction is attributed to the nonmetricity of a flat, symmetric, but not metric-compatible affine connection, have been a topic of growing interest in recent studies. Numerous works study the cosmology of symmetric teleparallel gravity assuming a flat Friedmann-Lemaître-Robertson-Walker metric, while working in the so-called ``coincident gauge'', further assuming that the connection coefficients vanish. However, little attention has been paid to the fact that both of these assumptions rely on the freedom to choose a particular coordinate system in order to simplify the metric or the connection, and that they may, in general, not be achieved simultaneously. Here we construct the most general symmetric teleparallel geometry obeying the conditions of homogeneity and isotropy, without making any assumptions on the properties of the coordinates, and present our results in both the usual cosmological coordinates and in the coincident gauge. We find that in general these coordinates do not agree, and that assuming both to hold simultaneously allows only for a very restricted class of geometries. For the general case, we derive the energy-momentum-hypermomentum conservation relations and the cosmological dynamics of selected symmetric teleparallel gravity theories. Our results show that the symmetric teleparallel connection in general contributes another scalar quantity into the cosmological dynamics, which decouples only for a specific class of theories from the dynamics of the metric, and only if the most simple geometry is chosen, in which both assumed coordinate systems agree. Most notably, the $f(Q)$ class of theories falls into this class.
\end{abstract}

\maketitle

%\tableofcontents

\section{Introduction}\label{sec:intro}
Recent observations in cosmology, such as the tension between different measurements of the Hubble parameter~\cite{Planck:2018vyg}, hint towards physics beyond the cosmological standard model, which describes the dynamics of the universe through general relativity, a cosmological constant $\Lambda$ and cold dark matter (CDM), hence being known as $\Lambda$CDM model. A plethora of modified theories of gravity has been developed, which offer potential explanations for these observations~\cite{DiValentino:2021izs,Saridakis:2021lqd}. While the most traditional class of modifications departs from the common formulation of general relativity using the metric and its Levi-Civita connection, thereby attributing gravity to the curvature of the latter, a large number of so-called teleparallel theories exists, which attribute gravity to the torsion or nonmetricity of a flat connection~\cite{BeltranJimenez:2019tjy}, or both of them~\cite{BeltranJimenez:2019odq,Boehmer:2021aji}. In this article we focus on the latter class of theories, more precisely known as symmetric teleparallel gravity theories, in which the employed connection is free of torsion and curvature. Starting from the symmetric teleparallel equivalent of general relativity (STEGR)~\cite{Nester:1998mp}, numerous theories of this class have been developed and studied~\cite{Adak:2005cd,Adak:2008gd,Mol:2014ooa,BeltranJimenez:2017tkd,BeltranJimenez:2018vdo,Jarv:2018bgs,Runkla:2018xrv,Adak:2018vzk,Hohmann:2018xnb,Hohmann:2018wxu,Jimenez:2019yyx,Koivisto:2019ggr,Flathmann:2020zyj,DAmbrosio:2020nqu}.

Since a major motivation for the study of modified gravity theories comes from cosmology, also the cosmological dynamics of teleparallel gravity theories has been subject of numerous studies. While the cosmology of metric teleparallel theories has been thoroughly discussed in the literature~\cite{Cai:2015emx,Bahamonde:2021gfp}, symmetric teleparallel cosmology has only recently received growing attention. Particular attention has been devoted to cosmology in the $f(Q)$ class of theories~\cite{BeltranJimenez:2019tme,Lazkoz:2019sjl,Lu:2019hra,Barros:2020bgg,Bajardi:2020fxh,Ayuso:2020dcu,Mandal:2020buf,Frusciante:2021sio,Mandal:2021bfu,Khyllep:2021pcu,Anagnostopoulos:2021ydo,Pradhan:2021qnz,Esposito:2021ect,Dimakis:2021gby,Atayde:2021ujc}, or theories which couple (pseudo-)scalar fields to the nonmetricity~\cite{Li:2020xjt,Hohmann:2020dgy}. An additional coupling to energy-momentum has been considered~\cite{Xu:2019sbp,Xu:2020yeg,Arora:2020tuk,Pradhan:2020brv,Arora:2020met,Arora:2020iva,Zia:2021vhr,Yang:2021fjy,Bhattacharjee:2021hwm,Arora:2021jik,Solanki:2021qni,Godani:2021mld,Arora:2021tuh,Haghani:2021fpx,Pradhan:2021dpf,Pati:2021ach,Rudra:2021ksp,Gadbail:2021kgd,Agrawal:2021rur}. These studies make use of the fact that around any point in spacetime there exist local coordinates in which the coefficients of the symmetric teleparallel connection vanish identically. This choice of coordinates, which is known as the \emph{coincident gauge}, allows replacing the symmetric teleparallel covariant derivative with the usual partial derivatives, hence leading to a simplification of various terms appearing in the field equations. In addition, the metric is assumed to exhibit cosmological symmetry, hence being homogeneous and isotropic, and thus chosen to be of the (in particular spatially flat) Friedmann-Lemaître-Robertson-Walker (FLRW) type.

Although the aforementioned choice of the coincident gauge and the FLRW metric appears canonical and leads to consistent dynamics for the theories whose cosmological dynamics has been studied, it attempts to employ the freedom to choose a coordinate system in order to achieve two goals at the same time, by imposing a particular form for both the metric and the symmetric teleparallel connection, each of which relies on the freedom to choose coordinates. However tempting, there is no reason to assume a priori that there exists a common coordinate system in which both quantities simultaneously take the desired forms. Demanding that the metric takes the FLRW form determines the space and time coordinates to be adapted to the foliation of the spacetime with spatial hypersurfaces, while the coincident gauge is unique up to an affine transformation of the coordinates. Hence, fixing the coordinates by asserting one of the aforementioned goals does not leave in general sufficient coordinate freedom in order to also achieve the other one. In particular, it obstructs considering FLRW cosmologies beyond the spatially flat case, which have been shown to require a covariant treatment instead~\cite{Zhao:2021zab}.

The aim of this article is to use the covariant approach to symmetric teleparallel gravity~\cite{Zhao:2021zab} and to study the most general class of homogeneous and isotropic symmetric teleparallel geometries, without any a priori assumptions on the compatibility of the coincident gauge and the cosmological coordinates imposed by the FLRW metric. The starting point of this analysis is given by the vector fields which generate rotations and translations on the spatial hypersurfaces, and under which we impose the metric and the connection to be invariant, hence following a similar approach as in metric teleparallel geometry~\cite{Hohmann:2019nat,Hohmann:2020zre,DAmbrosio:2021pnd}. Imposing this condition as the basis for the definition of cosmological symmetry is motivated both mathematically, as it agrees with the notion of symmetry derived from the underlying Cartan geometry~\cite{Hohmann:2015pva}, and physically: retaining a homogeneous and isotropic metric throughout the cosmic evolution requires that the terms appearing in the field equation governing its evolution obey the same symmetry. Since the metric and the connection are, in general, coupled in the field equations, imposing the cosmological symmetry also on the connection guarantees that its contribution to the field equations preserves this symmetry. We then use these homogeneous and isotropic geometries in order to derive the cosmological dynamics for a number of symmetric teleparallel gravity theories.

The outline of this article is as follows. In section~\ref{sec:stg}, we give a brief review of the essential ingredients of symmetric teleparallel gravity theories. We then determine the most general class of homogeneous and isotropic symmetric teleparallel geometries in section~\ref{sec:cosmogeom} by using the usual cosmological coordinates, in which the metric has the well-known FLRW form. In section~\ref{sec:coinc} we then perform a coordinate transformation to the coincident gauge, in which the connection coefficients vanish. The conservation of energy-momentum-hypermomentum in these cosmological backgrounds is discussed in section~\ref{sec:enmomhyp}. For the geometries and matter discussed in the aforementioned sections, we derive the cosmological field equations of a number of symmetric teleparallel gravity theories in section~\ref{sec:theories}, and display them in a form which is independent of the original choice of coordinates. We end with a conclusion in section~\ref{sec:conclusion}.

\section{Symmetric teleparallel gravity}\label{sec:stg}
We begin with a brief review of the key ingredients of symmetric teleparallel gravity. The dynamical fields are a metric \(g_{\mu\nu}\) of Lorentzian signature and an independent affine connection with coefficients \(\Gamma^{\rho}{}_{\mu\nu}\), defining a covariant derivative \(\nabla_{\mu}\), which is constrained by the conditions of vanishing curvature
\begin{equation}\label{eq:nocurv}
R^{\rho}{}_{\sigma\mu\nu} = \partial_{\mu}\Gamma^{\rho}{}_{\sigma\nu} - \partial_{\nu}\Gamma^{\rho}{}_{\sigma\mu} + \Gamma^{\rho}{}_{\lambda\mu}\Gamma^{\lambda}{}_{\sigma\nu} - \Gamma^{\rho}{}_{\lambda\nu}\Gamma^{\lambda}{}_{\sigma\mu} = 0\,,
\end{equation}
and vanishing torsion
\begin{equation}\label{eq:notors}
T^{\rho}{}_{\mu\nu} = \Gamma^{\rho}{}_{\nu\mu} - \Gamma^{\rho}{}_{\mu\nu} = 0\,,
\end{equation}
but in general non-vanishing nonmetricity
\begin{equation}
Q_{\rho\mu\nu} = \nabla_{\rho}g_{\mu\nu}\,.
\end{equation}
There are different, equivalent possibilities to implement these conditions, either by imposing them a priori and correspondingly restricting the variation of the action to obtain the field equations, or by including Lagrange multipliers~\cite{Hohmann:2021fpr}. Following the former approach, we consider a general symmetric teleparallel action of the form
\begin{equation}\label{eq:genaction}
S = S_{\text{g}}[g, \Gamma] + S_{\text{m}}[\chi, g, \Gamma]\,,
%+ \int(r_{\mu}{}^{\nu\rho\sigma}R^{\mu}{}_{\nu\rho\sigma} + t_{\mu}{}^{\nu\rho}T^{\mu}{}_{\nu\rho})
\end{equation}
where \(\chi\) denotes a generic set of matter fields. Note that we have included the independent connection also in the matter action, thus allowing also for matter which couples directly to this connection~\cite{Harko:2018gxr,Lobo:2019xwp}. It then follows that the variation of the action is generically of the form
\begin{equation}\label{eq:metricgravactvar}
\delta S_{\text{g}} = -\int_M\left(\frac{1}{2}W^{\mu\nu}\sqrt{-g}\delta g_{\mu\nu} + \tilde{Y}_{\mu}{}^{\nu\rho}\delta\Gamma^{\mu}{}_{\nu\rho}\right)\dd^4x
\end{equation}
for the gravitational part, as well as
\begin{equation}\label{eq:metricmatactvar}
\delta S_{\text{m}} = \int_M\left(\frac{1}{2}\Theta^{\mu\nu}\sqrt{-g}\delta g_{\mu\nu} + \tilde{H}_{\mu}{}^{\nu\rho}\delta\Gamma^{\mu}{}_{\nu\rho} + \tilde{\Omega}_I\delta\chi^I\right)\dd^4x
\end{equation}
for the matter part, where a tilde over a symbol indicates that the corresponding quantity is a tensor density. Keeping in mind that the variation of the connection must be of the form \(\delta\Gamma^{\mu}{}_{\nu\rho} = \nabla_{\rho}\nabla_{\nu}\zeta^{\mu}\) with a vector field \(\zeta^{\mu}\) in order to preserve the vanishing torsion and curvature, it follows that the field equations are given by~\cite{Hohmann:2021fpr}
\begin{equation}
W_{\mu\nu} = \Theta_{\mu\nu}\,, \quad
\nabla_{\nu}\nabla_{\rho}\tilde{Y}_{\mu}{}^{\nu\rho} = \nabla_{\nu}\nabla_{\rho}\tilde{H}_{\mu}{}^{\nu\rho}\,, \quad
\tilde{\Omega}_I = 0\,.
\end{equation}
Further, we demand that each of the terms \(S_{\text{g}}\) and \(S_{\text{m}}\) is separately invariant under diffeomorphisms. For the gravitational side, this implies the Bianchi identity
\begin{equation}
\sqrt{-g}\lc{\nabla}_{\nu}W_{\mu}{}^{\nu} = \nabla_{\nu}\nabla_{\rho}\tilde{Y}_{\mu}{}^{\nu\rho}\,,
\end{equation}
which is a geometric identity, i.e., it holds irrespective of the field equations for any symmetric teleparallel geometry. Here and in the remainder of this article \(\lc{\nabla}_{\mu}\) denotes the Levi-Civita covariant derivative defined by the metric, and we use a circle also to denote related quantities. In contrast, for the matter side we must impose the matter field equations \(\tilde{\Omega}_I = 0\) to be satisfied, in order to arrive at the energy-momentum-hypermomentum conservation law
\begin{equation}\label{eq:enmomhypcons}
\sqrt{-g}\lc{\nabla}_{\nu}\Theta_{\mu}{}^{\nu} = \nabla_{\nu}\nabla_{\rho}\tilde{H}_{\mu}{}^{\nu\rho}\,.
\end{equation}
Note that this relation is independent of the gravitational part \(S_{\text{g}}\) of the action and depends only on the matter part \(S_{\text{m}}\). Further, we see that imposing the metric field equation \(W_{\mu\nu} = \Theta_{\mu\nu}\), the connection field equation is automatically satisfied as a consequence of the Bianchi identities and the energy-momentum-hypermomentum conservation. We will make use of this fact throughout this article and display only the metric field equations for any theory we consider, while omitting the connection field equations, as they follow from the former.

\section{Homogeneous and isotropic symmetric teleparallel geometry}\label{sec:cosmogeom}
In order to derive the cosmological dynamics of symmetric teleparallel gravity theories outlined in the preceding section, we start by constructing the most general metric-affine geometry which obeys the cosmological symmetry, i.e., which is both homogeneous and isotropic. Hence, we start with a metric \(g_{\mu\nu}\) and an affine connection with coefficients \(\Gamma^{\mu}{}_{\nu\rho}\), and demand that their Lie derivatives~\cite{Yano:1957lda}
\begin{equation}\label{eq:metsymcondi}
(\mathcal{L}_{X}g)_{\mu\nu} = X^{\rho}\partial_{\rho}g_{\mu\nu} + \partial_{\mu}X^{\rho}g_{\rho\nu} + \partial_{\nu}X^{\rho}g_{\mu\rho}
\end{equation}
and
\begin{equation}\label{eq:affsymcondi}
(\mathcal{L}_{X}\Gamma)^{\mu}{}_{\nu\rho} = X^{\sigma}\partial_{\sigma}\Gamma^{\mu}{}_{\nu\rho} - \partial_{\sigma}X^{\mu}\Gamma^{\sigma}{}_{\nu\rho} + \partial_{\nu}X^{\sigma}\Gamma^{\mu}{}_{\sigma\rho} + \partial_{\rho}X^{\sigma}\Gamma^{\mu}{}_{\nu\sigma} + \partial_{\nu}\partial_{\rho}X^{\mu}\\
\end{equation}
vanish, where \(X^{\mu}\) is any of the three rotation generators
\begin{subequations}\label{eq:genrot}
\begin{align}
\varrho_1 &= \sin\varphi\partial_{\vartheta} + \frac{\cos\varphi}{\tan\vartheta}\partial_{\varphi}\,,\\
\varrho_2 &= -\cos\varphi\partial_{\vartheta} + \frac{\sin\varphi}{\tan\vartheta}\partial_{\varphi}\,,\\
\varrho_3 &= -\partial_{\varphi}\,,
\end{align}
\end{subequations}
or translation generators
\begin{subequations}\label{eq:gentra}
\begin{align}
\tau_1 &= \chi\sin\vartheta\cos\varphi\partial_r + \frac{\chi}{r}\cos\vartheta\cos\varphi\partial_{\vartheta} - \frac{\chi\sin\varphi}{r\sin\vartheta}\partial_{\varphi}\,,\\
\tau_2 &= \chi\sin\vartheta\sin\varphi\partial_r + \frac{\chi}{r}\cos\vartheta\sin\varphi\partial_{\vartheta} + \frac{\chi\cos\varphi}{r\sin\vartheta}\partial_{\varphi}\,,\\
\tau_3 &= \chi\cos\vartheta\partial_r - \frac{\chi}{r}\sin\vartheta\partial_{\vartheta}\,.
\end{align}
\end{subequations}
Here we used the abbreviation \(\chi = \sqrt{1 - kr^2}\), where \(k \in \mathbb{R}\) indicates the spatial curvature, and we work in spherical coordinates \((x^{\mu}) = (t, r, \vartheta, \varphi)\). The most general metric which satisfies these conditions is the well-known Robertson-Walker metric, whose non-vanishing components are given by
\begin{equation}\label{eq:metcosmo}
g_{tt} = -N^2\,, \quad
g_{rr} = \frac{A^2}{\chi^2}\,, \quad
g_{\vartheta\vartheta} = A^2r^2\,, \quad
g_{\varphi\varphi} = g_{\vartheta\vartheta}\sin^2\vartheta\,,
\end{equation}
where \(N = N(t)\) is the lapse function and \(A = A(t)\) is the scale factor. We include the former, as it will turn out useful to consider different choices for the time coordinate. Also for later use we decompose the metric in the form
\begin{equation}
g_{\mu\nu} = -n_{\mu}n_{\nu} + h_{\mu\nu}
\end{equation}
into the hypersurface conormal \(n_{\mu}\) and spatial metric \(h_{\mu\nu}\), whose non-vanishing components are given by
\begin{equation}
n_t = -N\,, \quad
h_{rr} = \frac{A^2}{\chi^2}\,, \quad
h_{\vartheta\vartheta} = A^2r^2\,, \quad
h_{\varphi\varphi} = h_{\vartheta\vartheta}\sin^2\vartheta\,.
\end{equation}
For the affine connection, the most general case which satisfies the conditions of cosmological symmetry is given by~\cite{Hohmann:2019fvf}
\begin{gather}
\Gamma^t{}_{tt} = K_1\,, \quad
\Gamma^{\vartheta}{}_{r\vartheta} = \Gamma^{\vartheta}{}_{\vartheta r} = \Gamma^{\varphi}{}_{r\varphi} = \Gamma^{\varphi}{}_{\varphi r} = \frac{1}{r}\,, \quad
\Gamma^{\varphi}{}_{\vartheta\varphi} = \Gamma^{\varphi}{}_{\varphi\vartheta} = \cot\vartheta\,, \quad
\Gamma^{\vartheta}{}_{\varphi\varphi} = -\sin\vartheta\cos\vartheta\,,\nonumber\\
\Gamma^t{}_{rr} = \frac{K_2}{\chi^2}\,, \quad
\Gamma^r{}_{\vartheta\vartheta} = -r\chi^2\,, \quad
\Gamma^r{}_{\varphi\varphi} = -r\chi^2\sin^2\vartheta\,, \quad
\Gamma^r{}_{\varphi\vartheta} = -\Gamma^r{}_{\vartheta\varphi} = K_5r^2\chi\sin\vartheta\,,\nonumber\\
\Gamma^t{}_{\vartheta\vartheta} = K_2r^2\,, \quad
\Gamma^r{}_{tr} = \Gamma^{\vartheta}{}_{t\vartheta} = \Gamma^{\varphi}{}_{t\varphi} = K_3\,, \quad
\Gamma^r{}_{rt} = \Gamma^{\vartheta}{}_{\vartheta t} = \Gamma^{\varphi}{}_{\varphi t} = K_4\,, \quad
\Gamma^r{}_{rr} = \frac{kr}{\chi^2}\,,\nonumber\\
\Gamma^t{}_{\varphi\varphi} = K_2r^2\sin^2\vartheta\,, \quad
\Gamma^{\vartheta}{}_{r\varphi} = -\Gamma^{\vartheta}{}_{\varphi r} = \frac{K_5\sin\vartheta}{\chi}\,, \quad
\Gamma^{\varphi}{}_{r\vartheta} = -\Gamma^{\varphi}{}_{\vartheta r} = -\frac{K_5}{\chi\sin\vartheta}\,,\label{eq:cosmoaffconn}
\end{gather}
where \(K_1(t), \ldots, K_5(t)\) are functions of time. In order to obtain a symmetric teleparallel geometry, we must further restrict this general cosmologically symmetric connection by imposing vanishing torsion and curvature. For the former we find that it can be written as
\begin{equation}
T^{\mu}{}_{\nu\rho} = 2\frac{K_4 - K_3}{N}h^{\mu}_{[\nu}n_{\rho]} + 2\frac{K_5}{A}\varepsilon^{\mu}{}_{\nu\rho}\,,
\end{equation}
where \(\varepsilon_{\mu\nu\rho} = n^{\sigma}\epsilon_{\sigma\mu\nu\rho}\) is the spatial part of the Levi-Civita tensor \(\epsilon_{\sigma\mu\nu\rho}\) of the metric \(g_{\mu\nu}\). We thus find that the torsion vanishes if and only if \(K_3 = K_4\) and \(K_5 = 0\) are satisfied. Using these conditions to eliminate \(K_4\) and \(K_5\), we find that we can write the curvature as
\begin{equation}
R^{\mu}{}_{\nu\rho\sigma} = 2\frac{K_3(K_3 - K_1) + \partial_tK_3}{N^2}n_{\nu}n_{[\rho}h^{\mu}_{\sigma]} + 2\frac{K_2(K_3 - K_1) - \partial_tK_2}{A^2}n^{\mu}n_{[\rho}h_{\sigma]\nu} + 2\frac{k + K_2K_3}{A^2}h^{\mu}_{[\rho}h_{\sigma]\nu}\,.
\end{equation}
The conditions on the remaining functions \(K_{1,2,3}\) for vanishing curvature thus take the simple form
\begin{equation}\label{eq:nocurvpar}
k + K_2K_3 = K_2(K_3 - K_1) - \partial_tK_2 = K_3(K_1 - K_3) - \partial_tK_3 = 0\,.
\end{equation}
In order to determine the most general solution to these conditions, it is useful to distinguish two cases:
\begin{enumerate}
\item
\(k = 0\): In this case \(K_2K_3 = 0\), so that at least one of these two functions must vanish. Depending on which of them vanishes, we can distinguish further the following subcases:
\begin{enumerate}
\item
\(K_2 = K_3 = 0\): In this case the equations involving \(K_1\) are solved identically, so that \(K_1\) remains arbitrary.
\item
\(K_2 \neq 0\): The equation involving \(\partial_tK_3\) is solved identically by \(K_3 = 0\), but now \(K_1\) is fixed to
\begin{equation}
K_1 = -\frac{\partial_tK_2}{K_2}\,,
\end{equation}
and \(K_2\) is the only remaining free function.
\item
\(K_3 \neq 0\): In this case, one similarly finds \(K_2 = 0\), while \(K_1\) is determined by
\begin{equation}
K_1 = K_3 + \frac{\partial_tK_3}{K_3}\,,
\end{equation}
so that only \(K_3\) remains arbitrary.
\end{enumerate}
\item
\(k \neq 0\): One has \(K_2K_3 = -k \neq 0\), fixing one of the two functions in terms of the other. Finally, \(K_1\) is fixed by
\begin{equation}
K_1 = K_3 + \frac{\partial_tK_3}{K_3} = K_3 - \frac{\partial_tK_2}{K_2}\,,
\end{equation}
where the last equality follows from \(\partial_t(K_2K_3) = 0\). Hence, also in this case only one function remains arbitrary. Note that a special case with \(\partial_tK_2 = \partial_tK_3 = 0\) has also been discussed in~\cite{Zhao:2021zab}.
\end{enumerate}
We see that in all of the aforementioned cases one of the functions in the connection coefficients remains arbitrary, while the remaining functions are fully determined. Simultaneously with our work, these branches have also been found and reported in~\cite{DAmbrosio:2021pnd}. For the study in the remainder of this article, it is helpful to choose a parametrization of this single function which will help to simplify the gravitational field equations of the various symmetric teleparallel gravity theories we discuss. The latter are constituted by the nonmetricity tensor, which takes the form
\begin{equation}\label{eq:cosmononmet}
Q_{\rho\mu\nu} = 2Q_1n_{\rho}n_{\mu}n_{\nu} + 2Q_2n_{\rho}h_{\mu\nu} + 2Q_3h_{\rho(\mu}n_{\nu)}
\end{equation}
for the torsion-free, but not necessarily flat connection parametrized by \(K_{1,2,3}\). Here the three expressions
\begin{equation}
Q_1 = \frac{\partial_tN}{N^2} - \frac{K_1}{N}\,, \quad
Q_2 = \frac{1}{N}\left(K_4 - \frac{\partial_tA}{A}\right)\,, \quad
Q_3 = \frac{K_3}{N} - \frac{K_2N}{A^2}\,,
\end{equation}
are scalar functions of the time coordinate \(t\), i.e., under a change \(t \mapsto \tilde{t}(t)\) of the time coordinate they transform trivially as
\begin{equation}
Q_i(t) = \tilde{Q}_i(\tilde{t}(t))\,,
\end{equation}
as does the scale factor \(A\). This stands in contrast to the original functions \(K_{1,2,3}\), as well as the lapse function \(N\), for which a non-trivial transformation law applies. It is therefore convenient to use the scalar functions \(Q_{1,2,3}\) in order to introduce a parametrization for the different branches of flat, symmetric connections we have found above. We start with the simpler case \(k = 0\):
\begin{enumerate}
\item
\(K_2 = K_3 = 0\): Choosing the remaining parameter function as
\begin{equation}\label{eq:parflat00}
K_1 = \frac{\partial_tN}{N} - KN
\end{equation}
in terms of a scalar function \(K = K(t)\) results in
\begin{equation}
Q_1 = K\,, \quad
Q_2 = -H\,, \quad
Q_3 = 0\,,
\end{equation}
where we introduced the Hubble parameter
\begin{equation}
H = \frac{\partial_tA}{NA} = \frac{\mathcal{L}_{n}A}{A} = \frac{\dot{A}}{A}\,.
\end{equation}
We use the Lie derivative here to emphasize again that this is a scalar function, and will henceforth denote it with a dot.
\item
\(K_2 \neq 0\): In this case choosing \(K_2\) also determines \(K_1\). It is convenient to choose
\begin{equation}\label{eq:parflat30}
K_2 = -\frac{KA^2}{N}\,, \quad
K_1 = \frac{\partial_tN}{N} - 2\frac{\partial_tA}{A} - \frac{\partial_tK}{K}\,,
\end{equation}
from which follows
\begin{equation}
Q_1 = 2H + \frac{\dot{K}}{K}\,, \quad
Q_2 = -H\,, \quad
Q_3 = K\,.
\end{equation}
\item
\(K_3 \neq 0\): In this case \(K_1\) is determined by \(K_3\), and we choose
\begin{equation}\label{eq:parflat20}
K_3 = KN\,, \quad
K_1 = KN + \frac{\partial_tN}{N} + \frac{\partial_tK}{K}\,,
\end{equation}
so that we obtain
\begin{equation}
Q_1 = -K - \frac{\dot{K}}{K}\,, \quad
Q_2 = K - H\,, \quad
Q_3 = K\,.
\end{equation}
\end{enumerate}
It is a remarkable fact that each of the three spatially flat branches given above can be obtained as a limit for \(k \to 0\) from the unique spatially curved branch \(k \neq 0\) by choosing different, but equivalent parametrizations for the latter. These can be chosen as follows:
\begin{enumerate}
\item
The most simple parametrization is given by
\begin{equation}\label{eq:parcurved}
K_1 = KN + \frac{\partial_tN}{N} + \frac{\partial_tK}{K}\,, \quad
K_2 = -\frac{k}{KN}\,, \quad
K_3 = KN\,,
\end{equation}
and results in
\begin{equation}
Q_1 = -K - \frac{\dot{K}}{K}\,, \quad
Q_2 = K - H\,, \quad
Q_3 = K + \frac{k}{KA^2}\,.
\end{equation}
In the limiting case \(k \to 0\), it reduces to the branch \(K_3 \neq 0\).
\item
Choosing instead the parametrization
\begin{equation}
K_1 = \frac{\partial_tN}{N} - 2\frac{\partial_tA}{A} - \frac{\partial_tK}{K} + \frac{kN}{KA^2}\,, \quad
K_2 = -\frac{KA^2}{N}\,, \quad
K_3 = \frac{kN}{KA^2}\,,
\end{equation}
and thus
\begin{equation}
Q_1 = 2H + \frac{\dot{K}}{K} - \frac{k}{KA^2}\,, \quad
Q_2 = \frac{k}{KA^2} - H\,, \quad
Q_3 = K + \frac{k}{KA^2}\,,
\end{equation}
one obtains the branch \(K_2 \neq 0\) in the limit \(k \to 0\).
\item
The third parametrization we present here is rather cumbersome, and we include it for completeness only. Setting
\begin{equation}
K_1 = \frac{\partial_tN}{N} - \frac{\partial_tA}{A} + \partial_t\tilde{K} + \sqrt{|k|}e^{\tilde{K}}\frac{N}{A}\, \quad
K_2 = -\sgn k\sqrt{|k|}e^{-\tilde{K}}\frac{A}{N}\,, \quad
K_3 = \sqrt{|k|}e^{\tilde{K}}\frac{N}{A}\,,
\end{equation}
one obtains
\begin{equation}
Q_1 = H - \dot{\tilde{K}} - \frac{\sqrt{|k|}}{A}e^{\tilde{K}}\,, \quad
Q_2 = \frac{\sqrt{|k|}}{A}e^{\tilde{K}} - H\,, \quad
Q_3 = \frac{\sqrt{|k|}}{A}\left(e^{\tilde{K}} + \sgn k\,e^{-\tilde{K}}\right)\,.
\end{equation}
In the limit \(k \to 0\), this yields the branch \(K_2 = K_3 = 0\) with
\begin{equation}
K = H - \dot{\tilde{K}}\,.
\end{equation}
\end{enumerate}
As can be seen from the form of \(K_2\) and \(K_3\) in the different parametrizations shown above, one can transform each of them into each other by a suitable redefinition of the parameter function \(K\), provided that \(k \neq 0\). For the remaining calculations in this article, we will therefore only use one of these parametrizations, and choose the first one given by the definition~\eqref{eq:parcurved}, as it is most simple. In the limit \(k \to 0\), however, the transformation of \(K\) becomes singular, so that the three spatially flat branches we presented are inequivalent and must be studied separately.

\section{Coordinate transformation and coincident gauge}\label{sec:coinc}
It follows from the conditions~\eqref{eq:nocurv} and~\eqref{eq:notors} of vanishing curvature and torsion that one can always find a local coordinate system in which the coefficients of the symmetric teleparallel connection vanish identically. This choice of coordinates, which is most often assumed in the literature on symmetric teleparallel gravity, is known as the coincident gauge. In this section we derive the coordinate transformation from the cosmological coordinates \((x^{\mu}) = (t, r, \vartheta, \varphi)\) to the coincident gauge \((\tilde{x}^{\mu})\) for the cosmologically symmetric connection branches we have derived in the previous section. The starting point of this calculation is an ansatz of the form
\begin{equation}\label{eq:coinccoord}
\tilde{x}^0 = \tilde{t}(t, r)\,, \quad
\tilde{x}^1 = \tilde{r}(t, r)\sin\vartheta\cos\varphi\,, \quad
\tilde{x}^2 = \tilde{r}(t, r)\sin\vartheta\sin\varphi\,, \quad
\tilde{x}^3 = \tilde{r}(t, r)\cos\vartheta\,,
\end{equation}
with two functions \(\tilde{t}\) and \(\tilde{r}\) of the coordinates \(t\) and \(r\) which are to be determined. From the condition that in the coincident gauge the connection coefficients vanish, \(\tilde{\Gamma}^{\mu}{}_{\nu\rho} \equiv 0\), follows that in the cosmological coordinates they take the form
\begin{equation}
\Gamma^{\mu}{}_{\nu\rho} = \frac{\partial x^{\mu}}{\partial\tilde{x}^{\sigma}}\frac{\partial^2\tilde{x}^{\sigma}}{\partial x^{\nu}\partial x^{\rho}}\,.
\end{equation}
Hence, we find a system of differential equations, from which \(\tilde{t}\) and \(\tilde{r}\) can be determined. From the coordinate transformation~\eqref{eq:coinccoord} we find the component equation
\begin{equation}
\frac{1}{r} = \Gamma^{\vartheta}{}_{r\vartheta} = \frac{\partial_r\tilde{r}}{\tilde{r}}\,,
\end{equation}
which is solved by the general ansatz
\begin{equation}
\tilde{r}(t, r) = R(t)r\,.
\end{equation}
We then use this solution and continue with the component equation
\begin{equation}\label{eq:compdist}
-r(1 - kr^2) = \Gamma^r{}_{\vartheta\vartheta} = \frac{rR\partial_t\tilde{t}}{r\partial_tR\partial_r\tilde{t} - R\partial_t\tilde{t}}\,.
\end{equation}
In order to solve this equation, we now distinguish between the different branches we have found in section~\ref{sec:cosmogeom}:
\begin{enumerate}
\item
We start with the case \(k \neq 0\). Then the equation~\eqref{eq:compdist} can be solved by making a separation ansatz of the form
\begin{equation}
\tilde{t}(t, r) = T(t)Z(r) \qquad \Rightarrow \qquad \partial_t\tilde{t} = Z\partial_tT\,, \quad \partial_r\tilde{t} = T\partial_rZ\,.
\end{equation}
With this ansatz we can separate the equation to obtain
\begin{equation}
\left(kr - \frac{1}{r}\right)\frac{\partial_rZ}{Z} = k\frac{R\partial_tT}{T\partial_tR}\,.
\end{equation}
Since the left hand side depends only on \(r\), while the right hand side depends only on \(t\), both must be constant. Denoting this constant by \(kc\) we find the solution
\begin{equation}
Z = a_1\left(\sqrt{1 - kr^2}\right)^c\,, \quad
T = a_2R^c\,,
\end{equation}
with integration constants \(a_1\) and \(a_2\), which can be merged into \(a = a_1a_2\) in the product \(\tilde{t}\). We further use this solution and find the component equation
\begin{equation}
\frac{kr}{1 - kr^2} = \Gamma^r{}_{rr} = \frac{kr}{1 - kr^2}[1 + (1 - c)kr^2]\,,
\end{equation}
which is solved by setting \(c = 1\). We have thus determined the coordinate transformation up to the free function \(R(t)\). Note that so far we have used only components of the general homogeneous and isotropic connection~\eqref{eq:cosmoaffconn} which are common to all branches, and we have not made use of the conditions of vanishing curvature and torsion yet. We now make use of the latter by replacing the free functions \(K_4\) and \(K_5\), so that we are left with only \(K_{1,2,3}\). This leads to the remaining equations
\begin{equation}
K_3 = \frac{\partial_tR}{R}\,, \quad
K_2 = -k\frac{R}{\partial_tR}\,, \quad
K_1 = \frac{\partial_t^2R}{\partial_tR}\,,
\end{equation}
which are consistent with the conditions~\eqref{eq:nocurvpar} of vanishing curvature. With the parametrization~\eqref{eq:parcurved}, we have
\begin{equation}\label{eq:krcurved}
K = \frac{\partial_tR}{RN} = \frac{\dot{R}}{R}\,,
\end{equation}
which can be integrated to obtain \(R(t)\) for any given function \(K(t)\). Absorbing the integration constant \(a\) found in the solution for \(\tilde{t}\) into a redefinition of \(R\) we thus have the coordinate transformation
\begin{equation}
\tilde{t} = R(t)\sqrt{1 - kr^2}\,, \quad
\tilde{r} = R(t)r\,.
\end{equation}

\item
In the case \(k = 0\), we find that \(\partial_t\tilde{t}\) cancels from the component equation~\eqref{eq:compdist}, and we are left with the equation
\begin{equation}
\partial_tR\partial_r\tilde{t} = 0\,,
\end{equation}
which can be solved either by \(\partial_tR = 0\) or \(\partial_r\tilde{t} = 0\), or both. We will find that these choices can be related to the three branches of spatially flat geometries as follows:
\begin{enumerate}
\item
Choosing \(\partial_r\tilde{t} = 0\) by setting \(\tilde{t}(t, r) = T(t)\) leads to the restriction \(K_2 = 0\) on the connection coefficients. Further, we find that \(T\) and \(R\) are related by
\begin{equation}
\partial_tR\partial_t^2T = \partial_t^2R\partial_tT\,,
\end{equation}
which is solved by
\begin{equation}
T(t) = T_0 + T_1R(t)\,.
\end{equation}
The remaining equations then read
\begin{equation}
K_3 = \frac{\partial_tR}{R}\,, \quad
K_1 = \frac{\partial_t^2R}{\partial_tR}\,,
\end{equation}
and finally lead to the same equation~\eqref{eq:krcurved} for \(R\) as in the spatially curved case when inserting the parametrization~\eqref{eq:parflat20} of the spatially flat branch \(K_3 \neq 0\). Finally, the integration constants \(T_0\) and \(T_1\) may be absorbed by a linear transformation of the time coordinate \(\tilde{t}\), which does not affect the coincident gauge, leading to the coordinate transformation
\begin{equation}
\tilde{t} = R(t)\,, \quad
\tilde{r} = R(t)r\,.
\end{equation}

\item
Similarly, choosing \(\partial_tR = 0\), and thus \(R(t) = R = \text{const.}\), we obtain the restriction \(K_3 = 0\). Further, we find the condition \(\partial_r\partial_t\tilde{t} = 0\), from which follows that \(\tilde{t}\) can be written as a sum
\begin{equation}
\tilde{t}(t, r) = T(t) + Z(r)\,.
\end{equation}
These unknown functions can be solved for using the remaining equation
\begin{equation}
K_2r = \frac{\partial_r\tilde{t}}{\partial_t\tilde{t}} = \frac{\partial_rR}{\partial_tT}\,.
\end{equation}
Separating this equation into parts depending on \(t\) and \(r\) only yields
\begin{equation}
K_2\partial_tT = \frac{\partial_rZ}{r}\,.
\end{equation}
Both sides must now be equal to the same constant, which we denote by \(c\). The right hand side is then solved by
\begin{equation}
Z(r) = \frac{c}{2}r^2 + Z_0\,,
\end{equation}
and the left hand side determines \(T\) from \(K_2\). With the parametrization~\eqref{eq:parflat30} it becomes
\begin{equation}
K = -\frac{cN}{A^2\partial_tT} = -\frac{c}{A^2\dot{T}}\,,
\end{equation}
so that finally \(T(t)\) is determined from the scalar function \(K(t)\). We can again fix the constants \(c = 1\) and \(Z_0 = 0\) by a linear transformation of \(\tilde{t}\), as well as \(R\) by a transformation of \(\tilde{r}\), and obtain
\begin{equation}
\tilde{t} = T(t) + \frac{1}{2}r^2\,, \quad
\tilde{r} = r\,.
\end{equation}

\item
Finally, combining both choices above to set \(\partial_tR = \partial_r\tilde{t} = 0\) leads to the condition \(K_2 = K_3 = 0\), corresponding to the last remaining branch. In this case the equation to determine \(T(t)\) is given by
\begin{equation}
\frac{\partial_t^2T}{\partial_tT} = K_1 = \frac{\partial_tN}{N} - KN\,,
\end{equation}
which can more succinctly be written as
\begin{equation}
\ddot{T} + K\dot{T} = 0\,,
\end{equation}
and can be integrated to obtain \(T(t)\) from \(K(t)\). Absorbing all integration constants using linear transformations as in the previous cases, the coordinate transformation finally takes the form
\begin{equation}
\tilde{t} = T(t)\,, \quad
\tilde{r} = r\,.
\end{equation}
\end{enumerate}
\end{enumerate}
We have thus found the coordinate transformation to the coincident gauge for all branches of flat, symmetric connections. We see that only for the spatially flat branch \(K_2 = K_3 = 0\) the coordinate transformation reduces to a reparametrization of the time coordinate, and so it is the only branch in which the metric retains the form~\eqref{eq:metcosmo}, with the transformation changing \(N\) and \(A\) only. Hence, choosing this form of the metric and assuming the coincident gauge \emph{a priori} inevitably fixes the branch \(K_2 = K_3 = 0\), and neglects the remaining branches.

It is instructive to calculate the symmetry generators~\eqref{eq:genrot} and~\eqref{eq:gentra} in the coincident gauge. From the form~\eqref{eq:affsymcondi} it follows \(\tilde{\partial}_{\mu}\tilde{\partial}_{\nu}\tilde{X}^{\rho} = 0\), and so their components must become affine functions of the coordinates. This is most obvious for the rotation generators, which read
\begin{equation}
\varrho_1 = \tilde{x}^3\tilde{\partial}_2 - \tilde{x}^2\tilde{\partial}_3\,, \quad
\varrho_2 = \tilde{x}^1\tilde{\partial}_3 - \tilde{x}^3\tilde{\partial}_1\,, \quad
\varrho_3 = \tilde{x}^2\tilde{\partial}_1 - \tilde{x}^1\tilde{\partial}_2
\end{equation}
in all branches, and thus reduce to the well-known Cartesian form. For the translation generators, the most simple Cartesian form
\begin{equation}
\tau_1 = \tilde{\partial}_1\,, \quad
\tau_2 = \tilde{\partial}_2\,, \quad
\tau_3 = \tilde{\partial}_3
\end{equation}
is obtained only in the spatially flat branch \(K_2 = K_3 = 0\). We further find the spatially flat cases
\begin{equation}
\tau_1 = \tilde{x}^0\tilde{\partial}_1\,, \quad
\tau_2 = \tilde{x}^0\tilde{\partial}_2\,, \quad
\tau_3 = \tilde{x}^0\tilde{\partial}_3
\end{equation}
in the branch \(K_3 \neq 0\) and
\begin{equation}
\tau_1 = \tilde{x}^1\tilde{\partial}_0 + \tilde{\partial}_1\,, \quad
\tau_2 = \tilde{x}^2\tilde{\partial}_0 + \tilde{\partial}_2\,, \quad
\tau_3 = \tilde{x}^3\tilde{\partial}_0 + \tilde{\partial}_3
\end{equation}
for \(K_2 \neq 0\). Finally, in the spatially curved case \(k \neq 0\) we have
\begin{equation}
\tau_1 = -k\tilde{x}^1\tilde{\partial}_0 + \tilde{x}^0\tilde{\partial}_1\,, \quad
\tau_2 = -k\tilde{x}^2\tilde{\partial}_0 + \tilde{x}^0\tilde{\partial}_2\,, \quad
\tau_3 = -k\tilde{x}^3\tilde{\partial}_0 + \tilde{x}^0\tilde{\partial}_3\,.
\end{equation}
One easily checks that they satisfy the same commutation relations as in their usual form.

While it is possible to work in the coincident gauge, it turns out to be simpler to use the cosmological coordinates, since it allows for a simpler form of the metric and its associated field equations, which will be the equations we study in the remainder of this article. Further, they allow for a simpler description of the energy-momentum-hypermomentum conservation, which we will derive in the following section.

\section{Cosmologically symmetric energy-momentum-hypermomentum}\label{sec:enmomhyp}
In order to model the cosmological dynamics of symmetric teleparallel gravity theories, one must impose the conditions of isotropy and homogeneity not only on the geometry as shown in the previous sections, but also on the matter variables defined from the variation~\eqref{eq:metricmatactvar}. For the energy-momentum tensor this leads to the well-known perfect fluid form
\begin{equation}
\Theta_{\mu\nu} = \rho n_{\mu}n_{\nu} + ph_{\mu\nu}
\end{equation}
defining the density \(\rho\) and pressure \(p\), while the hypermomentum is most generally given by a hyperfluid~\cite{Iosifidis:2020gth}
\begin{equation}
H_{\rho\mu\nu} = \phi h_{\mu\rho}n_{\nu} + \chi h_{\nu\rho}n_{\mu} + \psi h_{\mu\nu}n_{\rho} + \omega n_{\mu}n_{\nu}n_{\rho} - \zeta\varepsilon_{\rho\mu\nu}\,,
\end{equation}
where we introduced the density factor \(\tilde{H}_{\rho}{}^{\mu\nu} = H_{\rho}{}^{\mu\nu}\sqrt{-g}\). It is instructive to calculate the explicit form of the energy-momentum-hypermomentum conservation law~\eqref{eq:enmomhypcons} for the hyperfluid. For the left hand side we find the usual form
\begin{equation}
\lc{\nabla}_{\nu}\Theta_{\mu}{}^{\nu} = [\dot{\rho} + 3H(\rho + p)]n_{\mu}\,,
\end{equation}
which vanishes in the case of a perfect fluid, in which hypermomentum is absent. For the latter, we obtain the expression
\begin{multline}
\frac{\nabla_{\nu}\nabla_{\rho}(H_{\mu}{}^{\nu\rho}\sqrt{-g})}{\sqrt{-g}} = \big\{\ddot{\omega} + (6H + Q_1)\dot{\omega} + 3[\dot{H} + H(3H + Q_1)]\omega\\
+ 3(H + Q_2 - Q_3)[\dot{\psi} + (3H + 2Q_1 + 2Q_2)\psi] + 3(H + Q_2)[\dot{\phi} + \dot{\chi} + 3H(\phi + \chi)]\big\}n_{\mu}
\end{multline}
for the general cosmologically symmetric nonmetricity~\eqref{eq:cosmononmet}, where we have made use of the conditions~\eqref{eq:nocurvpar} of vanishing curvature in order to eliminate time derivatives \(\dot{Q}_2\) and \(\dot{Q}_3\), after replacing them with \(K_{1,2,3}\). To proceed further, one must consider the different branches of cosmologically symmetric geometries we found in section~\ref{sec:cosmogeom} separately in order to substitute the parameter functions \(Q_{1,2,3}\). The most simple case is given by \(K_2 = K_3 = 0\), which leads to \(Q_3 = 0\) and \(Q_2 = -H\), so that numerous terms cancel and we are left with
\begin{equation}
\frac{\nabla_{\nu}\nabla_{\rho}(H_{\mu}{}^{\nu\rho}\sqrt{-g})}{\sqrt{-g}} = \left\{\ddot{\omega} + (6H + K)\dot{\omega} + 3[\dot{H} + H(3H + K)]\omega\right\}n_{\mu}\,,
\end{equation}
depending on \(\omega\) only. Similarly, for \(K_2 \neq 0\) we find
\begin{equation}
\frac{\nabla_{\nu}\nabla_{\rho}(H_{\mu}{}^{\nu\rho}\sqrt{-g})}{\sqrt{-g}} = \left\{\ddot{\omega} + \left(8H + \frac{\dot{K}}{K}\right)\dot{\omega} + 3\left[\dot{H} + H\left(5H + \frac{\dot{K}}{K}\right)\right]\omega - 3K\left[\dot{\psi} + \left(5H + 2\frac{\dot{K}}{K}\right)\psi\right]\right\}n_{\mu}\,,
\end{equation}
while the branch \(K_3 \neq 0\) leads to
\begin{equation}
\frac{\nabla_{\nu}\nabla_{\rho}(H_{\mu}{}^{\nu\rho}\sqrt{-g})}{\sqrt{-g}} = \left\{\ddot{\omega} + \left(6H - K - \frac{\dot{K}}{K}\right)\dot{\omega} + 3\left[\dot{H} + H\left(3H - K - \frac{\dot{K}}{K}\right)\right]\omega + 3K[\dot{\phi} + \dot{\chi} + 3H(\phi + \chi)]\right\}n_{\mu}\,.
\end{equation}
Hence, we find that depending on the choice of the branch, either the hypermomentum component \(\psi\) or \(\phi + \chi\) contributes. Finally, in the curved case \(k \neq 0\) we have
\begin{multline}
\frac{\nabla_{\nu}\nabla_{\rho}(H_{\mu}{}^{\nu\rho}\sqrt{-g})}{\sqrt{-g}} = \Bigg\{\ddot{\omega} + \left(6H - K - \frac{\dot{K}}{K}\right)\dot{\omega} + 3\left[\dot{H} + H\left(3H - K - \frac{\dot{K}}{K}\right)\right]\omega\\
- 3\frac{k}{A^2K}\left[\dot{\psi} + \left(H - 2\frac{\dot{K}}{K}\right)\psi\right]+ 3K[\dot{\phi} + \dot{\chi} + 3H(\phi + \chi)]\Bigg\}n_{\mu}\,,
\end{multline}
and so both of the aforementioned hypermomentum components contribute.

We emphasize again that the relations we derived in this section hold independently of the choice of the gravity theory, and that they can be used, together with the metric field equation, in order to replace the field equation derived by variation of the total action with respect to the symmetric teleparallel connection. Hence, in the following section, in which we derive the cosmological field equations for a number of symmetric teleparallel gravity theories, we will restrict ourselves to displaying only the metric field equations. Even though the hypermomentum does not appear in these field equations, it contributes to the dynamics through the energy-momentum-hypermomentum conservation~\eqref{eq:enmomhypcons}, or equivalently, the connection field equations.

\section{Application and cosmological field equations}\label{sec:theories}
We will now use the most general homogeneous and isotropic symmetric teleparallel geometry derived in the section~\ref{sec:cosmogeom} and apply it to the field equations of a number of gravity theories. Here we make use of the fact that the field equation derived by variation with respect to the flat, symmetric connection is related to the metric field equations through the diffeomorphism invariance of the action, and thus not independent of the latter~\cite{Hohmann:2021fpr}. We will therefore display only the metric field equations. In particular, we discuss $f(Q)$ gravity in section~\ref{ssec:fq}, newer general relativity in section~\ref{ssec:ngr} and scalar-nonmetricity gravity in section~\ref{ssec:sng}.

\subsection{$f(Q)$ gravity}\label{ssec:fq}
As a first example we study the $f(Q)$ class of gravity theories, which is defined by the action~\cite{BeltranJimenez:2017tkd}
\begin{equation}
S_{\text{g}} = -\frac{1}{2\kappa^2}\int\dd^4x\sqrt{-g}f(Q)\,.
\end{equation}
Here \(f\) is a free function, which determines a particular theory within the general class, while \(Q\) is the nonmetricity scalar defined by
\begin{equation}\label{eq:nonmetscal}
Q = -Q_{\mu\nu\rho}P^{\mu\nu\rho} = \frac{1}{4}Q_{\mu\nu\rho}Q^{\mu\nu\rho} - \frac{1}{2}Q_{\mu\nu\rho}Q^{\mu\nu\rho} - \frac{1}{4}Q_{\rho\mu}{}^{\mu}Q^{\rho\nu}{}_{\nu} + \frac{1}{2}Q^{\mu}{}_{\mu\rho}Q^{\rho\nu}{}_{\nu}\,,
\end{equation}
where we introduced the nonmetricity conjugate
\begin{equation}\label{eq:nonmetconj}
P^{\rho\mu\nu} = -\frac{1}{4}Q^{\rho\mu\nu} + \frac{1}{2}Q^{(\mu\nu)\rho} + \frac{1}{4}g^{\mu\nu}(Q^{\rho\sigma}{}_{\sigma} - Q_{\sigma}{}^{\sigma\rho}) - \frac{1}{4}g^{\rho(\mu}Q^{\nu)\sigma}{}_{\sigma}\,.
\end{equation}
The nonmetricity scalar is related to the Ricci scalar of the Levi-Civita connection via the relation
\begin{equation}
\lc{R} = -Q - \lc{\nabla}_{\mu}(Q^{\mu\nu}{}_{\nu} - Q_{\nu}{}^{\nu\mu})\,.
\end{equation}
Hence, in the case \(f(Q) = Q\), the gravitational action reduces to the action of STEGR~\cite{Nester:1998mp}, which equals the Einstein-Hilbert action up to a boundary term. The metric field equations can most compactly be written as
\begin{equation}
\kappa^2\Theta^{\mu}{}_{\nu} = \frac{2}{\sqrt{-g}}\nabla_{\rho}\left(\sqrt{-g}P^{\rho\mu}{}_{\nu}f'\right) + P^{\mu\rho\sigma}Q_{\nu\rho\sigma}f' + \frac{1}{2}\delta^{\mu}_{\nu}f\,,
\end{equation}
where we denote derivatives of \(f\) with respect to \(Q\) by primes. To derive the cosmological dynamics of this class of theories, we insert the expressions we derived for the four branches of homogeneous and isotropic symmetric teleparallel geometries in section~\ref{sec:cosmogeom} into these field equations. The most simple case is given by the spatially flat branch \(K_2 = K_3 = 0\), for which we obtain
\begin{subequations}
\begin{align}
12H^2f' - f &= 2\kappa^2\rho\,,\\
-48H^2\dot{H}f'' - 4(\dot{H} + 3H^2)f' + f &= 2\kappa^2p\,.
\end{align}
\end{subequations}
It is remarkable that in this case the scalar \(K\) which determines the symmetric teleparallel connection fully decouples from the metric field equations, so that the latter reduce to the equations derived by using the coincident gauge~\cite{BeltranJimenez:2017tkd}; however, it still enters the dynamics through the energy-momentum-hypermomentum conservation discussed in section~\ref{sec:enmomhyp}, unless the hypermomentum component \(\omega\) vanishes. Moreover, we see that the first equation turns into a constraint equation, generalizing the well-known Friedmann constraint equation. However, to see that this is only a particular property of the chosen geometry branch, and not a generic feature of symmetric teleparallel gravity theories, we also derive the dynamics for the other spatially flat branches. In the case \(K_2 \neq 0\) we have
\begin{subequations}
\begin{align}
9K[\ddot{K} + 3H\dot{K} + (3K + 4H)\dot{H}]f'' + 3(\dot{K} + 3HK + 4H^2)f' - f &= 2\kappa^2\rho\,,\\
-3(K + 4H)[\ddot{K} + 3H\dot{K} + (3K + 4H)\dot{H}]f'' - [3\dot{K} + 4\dot{H} + 3H(3K + 4H)]f' + f &= 2\kappa^2p\,,
\end{align}
\end{subequations}
while for the branch \(K_3 \neq 0\) we find
\begin{subequations}
\begin{align}
-9K[\ddot{K} + 3H\dot{K} + (3K - 4H)\dot{H}]f'' - 3(\dot{K} + 3HK - 4H^2)f' - f &= 2\kappa^2\rho\,,\\
-3(3K - 4H)[\ddot{K} + 3H\dot{K} + (3K - 4H)\dot{H}]f'' + [3\dot{K} - 4\dot{H} + 3H(3K - 4H)]f' + f &= 2\kappa^2p\,.
\end{align}
\end{subequations}
We see that now \(K\) enters as a new dynamical field. Finally, for the spatially curved (\(k \neq 0\)) branch we obtain the rather lengthy equations
\begin{subequations}
\begin{align}
-9\Bigg\{\frac{2k}{A^2}\left[\frac{\ddot{K}}{K} - \frac{\dot{K}^2}{K^2} + \frac{H\dot{K}}{K} - 2\frac{H\dot{H}}{K} + \dot{H} - 2HK + H^2\right] + \frac{k^2}{A^4K^2}\left[\frac{\ddot{K}}{K} - 2\frac{\dot{K}^2}{K^2} - \frac{H\dot{K}}{K} - \dot{H} - 4HK + 2H^2\right] &\nonumber\\
\phantom{0}+ K[\ddot{K} + 3H\dot{K} + (3K - 4H)\dot{H}]\Bigg\}f'' - 3\left[\frac{k}{A^2K^2}(\dot{K} - HK) + \dot{K} + 3HK - 4H^2\right]f' - f &= 2\kappa^2\rho\,,\\
\Bigg\{\frac{6k}{A^2}\left[\left(2\frac{H}{K} - 1\right)\frac{\ddot{K}}{K} + \left(3 - 4\frac{H}{K}\right)\left(\frac{\dot{K}^2}{K^2} + \dot{H} + 2HK - H^2\right) + \left(3 - 2\frac{H}{K}\right)\frac{H\dot{K}}{K} \right] &\nonumber\\
\phantom{0}+ \frac{3k^2}{A^4K^2}\left[\frac{\ddot{K}}{K} - 2\frac{\dot{K}^2}{K^2} - \frac{H\dot{K}}{K} - \dot{H} - 4HK + 2H^2\right] - 3(3K - 4H)[\ddot{K} + 3H\dot{K} + (3K - 4H)\dot{H}]\Bigg\}f'' &\nonumber\\
\phantom{0}+ \left[\frac{k}{A^2}\left(4 - 3\frac{H}{K} + 3\frac{\dot{K}}{K^2}\right) + 3\dot{K} - 4\dot{H} + 3H(3K - 4H)\right]f' + f &= 2\kappa^2p\,.
\end{align}
\end{subequations}
where we have used the parametrization~\eqref{eq:parcurved}, so that they reduce to the case \(K_3 \neq 0\) in the limit \(k \to 0\). In order to show that all cosmological field equations shown above indeed reduce to the case of general relativity for the parameter function \(f(Q) = Q\), it is helpful to display the value of \(Q\) for all branches, which follow from the general expression
\begin{equation}
Q = 3(2Q_2^2 + Q_1Q_3 - Q_2Q_3)\,.
\end{equation}
In the case \(K_2 = K_3 = 0\) we find the known result
\begin{equation}
Q = 6H^2\,,
\end{equation}
while for \(K_2 \neq 0\) we have
\begin{equation}
Q = 3(2H^2 + 3HK + \dot{K})\,,
\end{equation}
and \(K_3 \neq 0\) yields
\begin{equation}
Q = 3(2H^2 - 3HK - \dot{K})\,;
\end{equation}
finally, the spatially curved branch yields
\begin{equation}
Q = 3\left(2H^2 - 3HK - \dot{K} + k\frac{HK - 2K^2 - \dot{K}}{A^2K^2}\right)\,.
\end{equation}
Inserting these values for \(f = Q\) in the cosmological field equations, together with \(f' = 1\) and \(f'' = 0\), we find that the terms involving \(K\) indeed cancel, and we are left with the cosmological dynamics of general relativity.

We finally remark that the cosmological field equations shown in this section have simultaneously been derived in~\cite{DAmbrosio:2021pnd}, where also an exact solution for a power law model \(f(Q) \sim Q^{\alpha}\) has been studied.

\subsection{Newer General Relativity}\label{ssec:ngr}
The second theory whose cosmological dynamics we derive is newer general relativity, whose action we write in the form~\cite{BeltranJimenez:2017tkd}
\begin{equation}
S_{\text{g}} = -\frac{1}{2\kappa^2}\int\dd^4x\sqrt{-g}(c_1Q^{\mu\nu\rho}Q_{\mu\nu\rho} + c_2Q^{\mu\nu\rho}Q_{\rho\mu\nu} + c_3Q^{\rho\mu}{}_{\mu}Q_{\rho\nu}{}^{\nu} + c_4Q^{\mu}{}_{\mu\rho}Q_{\nu}{}^{\nu\rho} + c_5Q^{\mu}{}_{\mu\rho}Q^{\rho\nu}{}_{\nu})\,,
\end{equation}
where \(c_1, \ldots, c_5\) are constants. If these constants are chosen to take the values
\begin{equation}\label{eq:ngrstegr}
c_1 = -\frac{1}{4}\,, \quad
c_2 = \frac{1}{2}\,, \quad
c_3 = \frac{1}{4}\,, \quad
c_4 = 0\,, \quad
c_5 = -\frac{1}{2}\,,
\end{equation}
we find that the theory reduces to STEGR. The field equations obtained by variation with respect to the metric are given by
\begin{equation}\label{eq:ngrfield}
\begin{split}
\kappa^2\Theta_{\mu\nu} &= -2\lc{\nabla}_{\rho}\left[c_1Q^{\rho}{}_{\mu\nu} + c_2Q_{(\mu\nu)}{}^{\rho} + c_3Q^{\rho\sigma}{}_{\sigma}g_{\mu\nu} + c_4Q^{\sigma}{}_{\sigma(\mu}\delta_{\nu)}^{\rho} + \frac{1}{2}c_5\left(Q_{\sigma}{}^{\sigma\rho}g_{\mu\nu} + \delta^{\rho}_{(\mu}Q_{\nu)\sigma}{}^{\sigma}\right)\right]\\
&\phantom{=}+ \frac{1}{2}(c_1Q^{\rho\sigma\tau}Q_{\rho\sigma\tau} + c_2Q^{\rho\sigma\tau}Q_{\tau\rho\sigma} + c_3Q^{\tau\rho}{}_{\rho}Q_{\tau\sigma}{}^{\sigma} + c_4Q^{\rho}{}_{\rho\tau}Q_{\sigma}{}^{\sigma\tau} + c_5Q^{\rho}{}_{\rho\tau}Q^{\tau\sigma}{}_{\sigma})g_{\mu\nu} - c_3Q_{\mu\rho}{}^{\rho}Q_{\nu\sigma}{}^{\sigma}\\
&\phantom{=}+ c_1\left(2Q^{\rho\sigma}{}_{\mu}Q_{\sigma\rho\nu} - Q_{\mu}{}^{\rho\sigma}Q_{\nu\rho\sigma} - 2Q^{\rho\sigma}{}_{(\mu}Q_{\nu)\rho\sigma}\right) + c_2\left(Q^{\rho\sigma}{}_{\mu}Q_{\rho\sigma\nu} - Q_{\mu}{}^{\rho\sigma}Q_{\nu\rho\sigma} - Q^{\rho\sigma}{}_{(\mu}Q_{\nu)\rho\sigma}\right)\\
&\phantom{=}+ c_4\left[Q_{\rho}{}^{\rho\sigma}\left(Q_{\sigma\mu\nu} - 2Q_{(\mu\nu)\sigma}\right) + Q^{\rho}{}_{\rho\mu}Q^{\sigma}{}_{\sigma\nu} - Q^{\rho}{}_{\rho(\mu}Q_{\nu)\sigma}{}^{\sigma}\right] + \frac{1}{2}c_5\left[Q^{\rho\sigma}{}_{\sigma}\left(Q_{\rho\mu\nu} - 2Q_{(\mu\nu)\rho}\right) - Q_{\mu\rho}{}^{\rho}Q_{\nu\sigma}{}^{\sigma}\right]\,.
\end{split}
\end{equation}
In order to display the cosmological field equations for the homogeneous and isotropic geometries listed in section~\ref{sec:cosmogeom}, it is helpful to replace the original constants in the gravitational action by the linear combinations
\begin{equation}
a_1 = 2(c_1 + 3c_3)\,, \quad
a_2 = 2(2c_3 + c_5)\,, \quad
a_3 = 2(c_1 + c_2 + c_3 + c_4 + c_5)\,, \quad
a_4 = 2(c_2 - c_4 + c_5)\,,
\end{equation}
which turn out to be the only linear combinations which enter into the resulting equations, i.e., the cosmological dynamics turn out to be unchanged if we change the theory by the replacement
\begin{equation}\label{eq:ngrparshift}
c_1 \mapsto c_1 - 3\epsilon\,, \quad
c_2 \mapsto c_2 + 3\epsilon\,, \quad
c_3 \mapsto c_3 + \epsilon\,, \quad
c_4 \mapsto c_4 + \epsilon\,, \quad
c_5 \mapsto c_5 - 2\epsilon
\end{equation}
of the parameters with a free constant \(\epsilon\). The new linear combinations are chosen such that in the STEGR case~\eqref{eq:ngrstegr} they take the values \(a_1 = 1\) and \(a_2 = a_3 = a_4 = 0\). Using these constants, we find that the cosmological field equations for the branch \(K_2 = K_3 = 0\) read
\begin{subequations}
\begin{align}
2a_3\dot{K} + 3a_2\dot{H} + a_3K^2 + 3(a_2 + 2a_3)HK + 3(a_1 + 3a_2)H^2 &= \kappa^2\rho\,,\\
-a_2\dot{K} - 2a_1\dot{H} + a_3K^2 - 3a_1H^2 &= \kappa^2p\,.
\end{align}
\end{subequations}
Taking a closer look at these equations, we see that they depend only on the three constants \(a_{1,2,3}\), and that the scalar \(K\) related to the symmetric teleparallel connection decouples for \(a_2 = a_3 = 0\). In the latter case, we find that the cosmological dynamics become identical to those of general relativity, if one further chooses the normalization \(a_1 = 1\). Since these are three linearly independent conditions on the five initial parameters \(c_1, \ldots, c_5\), it follows that there exists a two-parameter family of theories which exhibits this property. This observation is different for the remaining branches. With the branch \(K_2 \neq 0\) we obtain the equations
\begin{subequations}
\begin{align}
2a_3\frac{\ddot{K}}{K} - a_3\frac{\dot{K}^2}{K^2} + (3a_2 + 10a_3)\frac{H\dot{K}}{K} + (3a_2 + 4a_3)\dot{H} + \frac{9}{4}(a_2 - 2a_3 + a_4)K^2 + 3a_4HK + (3a_1 + 15a_2 + 16a_3)H^2 &= \kappa^2\rho\,,\\
-a_2\frac{\ddot{K}}{K} + (a_2 + a_3)\frac{\dot{K}^2}{K^2} + 4a_3\frac{H\dot{K}}{K} + a_4\dot{K} - 2(a_1 + a_2)\dot{H} + \frac{a_2 - 2a_3 + a_4}{4}K^2 + 2a_4HK + (4a_3 - 3a_1)H^2 &= \kappa^2p\,.
\end{align}
\end{subequations}
while for \(K_3 \neq 0\) we find
\begin{subequations}
\begin{align}
-2a_3\frac{\ddot{K}}{K} + 3a_3\frac{\dot{K}^2}{K^2} - 3(a_2 + 2a_3)\frac{H\dot{K}}{K} + 3a_2\dot{H} + \frac{21a_2 + 22a_3 - 3a_4}{4}K^2 - 3(6a_2 + 4a_3 - a_4)HK + 3(a_1 + 3a_2)H^2 &= \kappa^2\rho\,,\\
a_2\frac{\ddot{K}}{K} + (a_3 - a_2)\frac{\dot{K}^2}{K^2} + (6a_2 + 4a_3 - a_4)\dot{K} - 2a_1\dot{H} + \frac{21a_2 + 22a_3 - 3a_4}{4}K^2 - 3a_1H^2 &= \kappa^2p\,.
\end{align}
\end{subequations}
We see that in this case also \(a_4\) appears in the dynamical equations. For theories satisfying \(a_2 = a_3 = a_4\) we find that \(K\) decouples, after which the cosmological field equations reduce to those of general relativity. Together with the normalization condition \(a_1 = 1\) this leaves a one-parameter family of theories whose cosmological dynamics agrees with general relativity for these two branches, with the single free parameter given by the transformation~\eqref{eq:ngrparshift}. Finally, coming to the spatially curved case \(k \neq 0\) we find the cosmological field equations
\begin{subequations}
\begin{align}
-2a_3\frac{\ddot{K}}{K} + 3a_3\frac{\dot{K}^2}{K^2} - 3(a_2 + 2a_3)\frac{H\dot{K}}{K} + 3a_2\dot{H} + \frac{21a_2 + 22a_3 - 3a_4}{4}K^2 - 3(6a_2 + 4a_3 - a_4)HK &\nonumber\\
\phantom{0}+ 3(a_1 + 3a_2)H^2 + 3a_4k\frac{H}{A^2K} + \frac{3}{2}k\frac{2a_1 - a_2 - 4a_3 + a_4}{A^2} + \frac{9}{4}k^2\frac{a_2 - 2a_3 + a_4}{A^4K^2} &= \kappa^2\rho\,,\\
a_2\frac{\ddot{K}}{K} + (a_3 - a_2)\frac{\dot{K}^2}{K^2} + (6a_2 + 4a_3 - a_4)\dot{K} - 2a_1\dot{H} + \frac{21a_2 + 22a_3 - 3a_4}{4}K^2 - 3a_1H^2 &\nonumber\\
\phantom{0}- a_4k\frac{\dot{K}}{A^2K^2}- \frac{1}{2}k\frac{2a_1 - a_2 - 4a_3 + a_4}{A^2} + \frac{1}{4}k^2\frac{a_2 - 2a_3 + a_4}{A^4K^2} &= \kappa^2p\,.
\end{align}
\end{subequations}
We find that the behavior qualitatively agrees with the previous two branches and that \(K\) decouples if \(a_2 = a_3 = a_4 = 0\), leading to cosmological dynamics identical to general relativity for a spatially curved FLRW metric.

\subsection{Scalar-nonmetricity gravity}\label{ssec:sng}
As a last example, we consider a class of scalar-nonmetricity theories given by the gravitational action
\begin{equation}
S_{\text{g}} = \frac{1}{2\kappa^2}\int\left[-\mathcal{A}(\phi)Q + 2\mathcal{B}(\phi)X + 2\mathcal{C}(\phi)Y + 2\mathcal{D}(\phi)Z - 2\mathcal{V}(\phi)\right]\,,
\end{equation}
which generalizes the class of theories presented in~\cite{Jarv:2018bgs,Runkla:2018xrv}. Here \(Q\) denotes the nonmetricity scalar~\eqref{eq:nonmetscal}, and we further introduced the terms
\begin{equation}
X = -\frac{1}{2}g^{\mu\nu}\partial_{\mu}\phi\partial_{\nu}\phi\,, \quad
Y = Q^{\mu\nu}{}_{\nu}\partial_{\mu}\phi\,, \quad
Z = Q_{\nu}{}^{\nu\mu}\partial_{\mu}\phi\,.
\end{equation}
The general form of the action contains the free functions \(\mathcal{A}, \mathcal{B}, \mathcal{C}, \mathcal{D}, \mathcal{V}\) of the scalar field \(\phi\), whose choice selects a particular theory from the general class. The field equations can most conveniently be expressed using the nonmetricity conjugate~\eqref{eq:nonmetconj} and read
\begin{multline}
\kappa^2\Theta_{\mu\nu} = \frac{2}{\sqrt{-g}}\nabla_{\rho}\left(\sqrt{-g}\mathcal{A}P^{\rho\sigma}{}_{\nu}\right)g_{\mu\sigma} + \mathcal{A}P_{\mu}{}^{\rho\sigma}Q_{\nu\rho\sigma} + \frac{\mathcal{A}}{2}g_{\mu\nu}Q + \left(\frac{\mathcal{B}}{2} + 2\mathcal{C}'\right)\lc{\nabla}^{\rho}\phi\lc{\nabla}_{\rho}\phi g_{\mu\nu} - (\mathcal{B} - 2\mathcal{D}')\lc{\nabla}_{\mu}\phi\lc{\nabla}_{\nu}\phi\\
+ 2\mathcal{C}\lc{\square}\phi g_{\mu\nu} + 2\mathcal{D}\lc{\nabla}_{\mu}\lc{\nabla}_{\nu}\phi + (2\mathcal{C} + \mathcal{D})\lc{\nabla}_{(\mu}\phi Q_{\nu)\rho}{}^{\rho} - \mathcal{C}\lc{\nabla}_{\rho}\phi Q^{\rho\sigma}{}_{\sigma}g_{\mu\nu} + \mathcal{D}\lc{\nabla}^{\rho}\phi(2Q_{(\mu\nu)\rho} - Q_{\rho\mu\nu} - Q^{\sigma}{}_{\sigma\rho}g_{\mu\nu})  + \mathcal{V}g_{\mu\nu}
\end{multline}
for the metric, as well as
\begin{equation}
\mathcal{A}'Q + 2\mathcal{B}\lc{\square}\phi + \mathcal{B}'\lc{\nabla}_{\mu}\phi\lc{\nabla}^{\mu}\phi - 2\mathcal{C}\lc{\nabla}_{\mu}Q^{\mu\nu}{}_{\nu} - 2\mathcal{D}\lc{\nabla}_{\mu}Q_{\nu}{}^{\nu\mu} - 2\mathcal{V}' = 0
\end{equation}
for the scalar field. We then insert the geometries we determined in section~\ref{sec:cosmogeom}. For the branch \(K_2 = K_3 = 0\) we obtain the cosmological field equations
\begin{subequations}
\begin{align}
3\mathcal{A}H^2 + 2(\mathcal{C} + \mathcal{D})K\dot{\phi} + 6(2\mathcal{C} + \mathcal{D})H\dot{\phi} - \frac{\mathcal{B} - 4(\mathcal{C}' + \mathcal{D}')}{2}\dot{\phi}^2 + 2(\mathcal{C} + \mathcal{D})\ddot{\phi} - \mathcal{V} &= \kappa^2\rho\,,\\
-\mathcal{A}(2\dot{H} + 3H^2) + 2(\mathcal{C} + \mathcal{D})K\dot{\phi} - 2\mathcal{A}'H\dot{\phi} - \frac{\mathcal{B} + 4\mathcal{C}'}{2}\dot{\phi}^2 - 2\mathcal{C}\ddot{\phi} + \mathcal{V} &= \kappa^2p\,,\\
12\mathcal{C}\dot{H} + 4(\mathcal{C} + \mathcal{D})\dot{K} + 6(6\mathcal{C} - \mathcal{A}')H^2 + 12(\mathcal{C} + \mathcal{D})KH - 6\mathcal{B}H\dot{\phi} - \mathcal{B}'\dot{\phi}^2 - 2\mathcal{B}\ddot{\phi} - 2\mathcal{V}' &= 0\,.
\end{align}
\end{subequations}
Again it is remarkable that there exists a special class of theories, now given by \(\mathcal{C} + \mathcal{D} = 0\), in which the scalar \(K\) from the symmetric teleparallel connection decouples. In this case the first equation becomes the usual Friedmann constraint equation. This class of theories also stands out in the remaining branches. For \(K_2 \neq 0\) we find the dynamical equations
\begin{subequations}
\begin{align}
3\mathcal{A}H^2 + 2(\mathcal{C} + \mathcal{D})\frac{\dot{K}\dot{\phi}}{K} + \frac{3}{2}(\mathcal{A}' + 2\mathcal{D})K\dot{\phi} + 2(8\mathcal{C} + 5\mathcal{D})H\dot{\phi} - \frac{\mathcal{B} - 4(\mathcal{C}' + \mathcal{D}')}{2}\dot{\phi}^2 + 2(\mathcal{C} + \mathcal{D})\ddot{\phi} - \mathcal{V} &= \kappa^2\rho\,,\\
-\mathcal{A}(2\dot{H} + 3H^2) + 2(\mathcal{C} + \mathcal{D})\frac{\dot{K}\dot{\phi}}{K} - \frac{1}{2}(\mathcal{A}' + 2\mathcal{D})K\dot{\phi} + (4\mathcal{C} + 4\mathcal{D} - 2\mathcal{A}')H\dot{\phi} - \frac{\mathcal{B} + 4\mathcal{C}'}{2}\dot{\phi}^2 - 2\mathcal{C}\ddot{\phi} + \mathcal{V} &= \kappa^2p\,,\\
4(\mathcal{C} + \mathcal{D})\frac{K\ddot{K} - \dot{K}^2 + 3HK\dot{K}}{K^2} - 3(\mathcal{A}' + 2\mathcal{D})(\dot{K} + 3HK) + 4(5\mathcal{C} + 2\mathcal{D})\dot{H} + 6(10\mathcal{C} + 4\mathcal{D} - \mathcal{A}')H^2 &\nonumber\\
\phantom{0}- 6\mathcal{B}H\dot{\phi} - \mathcal{B}'\dot{\phi}^2 - 2\mathcal{B}\ddot{\phi} - 2\mathcal{V}' &= 0\,,
\end{align}
\end{subequations}
while in the case \(K_3 \neq 0\) we have
\begin{subequations}
\begin{align}
3\mathcal{A}H^2 - 2(\mathcal{C} + \mathcal{D})\frac{\dot{K}\dot{\phi}}{K} + \frac{3\mathcal{A}' - 16\mathcal{C} - 10\mathcal{D}}{2}K\dot{\phi} + 6(2\mathcal{C} + \mathcal{D})H\dot{\phi} - \frac{\mathcal{B} - 4(\mathcal{C}' + \mathcal{D}')}{2}\dot{\phi}^2 + 2(\mathcal{C} + \mathcal{D})\ddot{\phi} - \mathcal{V} &= \kappa^2\rho\,,\\
-\mathcal{A}(2\dot{H} + 3H^2) - 2(\mathcal{C} + \mathcal{D})\frac{\dot{K}\dot{\phi}}{K} + \frac{1}{2}(3\mathcal{A}' - 16\mathcal{C} - 10\mathcal{D})K\dot{\phi} - 2\mathcal{A}'H\dot{\phi} - \frac{\mathcal{B} + 4\mathcal{C}'}{2}\dot{\phi}^2 - 2\mathcal{C}\ddot{\phi} + \mathcal{V} &= \kappa^2p\,,\\
4(\mathcal{C} + \mathcal{D})\frac{\dot{K}^2 - K\ddot{K} - 3HK\dot{K}}{K^2} + (3\mathcal{A}' - 16\mathcal{C} - 10\mathcal{D})(\dot{K} + 3HK) + 12\mathcal{C}\dot{H} + 6(6\mathcal{C} - \mathcal{A}')H^2 &\nonumber\\
\phantom{0}- 6\mathcal{B}H\dot{\phi} - \mathcal{B}'\dot{\phi}^2 - 2\mathcal{B}\ddot{\phi} - 2\mathcal{V}' &= 0\,.
\end{align}
\end{subequations}
Note that the second derivative \(\ddot{K}\) enters the field equations only for theories with \(\mathcal{C} + \mathcal{D} \neq 0\). This behavior can also be observed in the spatially curved branch \(k \neq 0\), which leads to the cosmological field equations
\begin{subequations}
\begin{align}
3\mathcal{A}H^2 - 2(\mathcal{C} + \mathcal{D})\frac{\dot{K}\dot{\phi}}{K} + \frac{3\mathcal{A}' - 16\mathcal{C} - 10\mathcal{D}}{2}K\dot{\phi} + 6(2\mathcal{C} + \mathcal{D})H\dot{\phi} &\nonumber\\
\phantom{0}- \frac{\mathcal{B} - 4(\mathcal{C}' + \mathcal{D}')}{2}\dot{\phi}^2 + 2(\mathcal{C} + \mathcal{D})\ddot{\phi} - \mathcal{V} + \frac{3k}{2A^2K}[2\mathcal{A}K + (\mathcal{A}' + 2\mathcal{D})\dot{\phi}] &= \kappa^2\rho\,,\\
-\mathcal{A}(2\dot{H} + 3H^2) - 2(\mathcal{C} + \mathcal{D})\frac{\dot{K}\dot{\phi}}{K} + \frac{1}{2}(3\mathcal{A}' - 16\mathcal{C} - 10\mathcal{D})K\dot{\phi} - 2\mathcal{A}'H\dot{\phi} &\nonumber\\
\phantom{0}- \frac{\mathcal{B} + 4\mathcal{C}'}{2}\dot{\phi}^2 - 2\mathcal{C}\ddot{\phi} + \mathcal{V} - \frac{k}{2A^2K}[2\mathcal{A}K + (\mathcal{A}' + 2\mathcal{D})\dot{\phi}] &= \kappa^2p\,,\\
4(\mathcal{C} + \mathcal{D})\frac{\dot{K}^2 - K\ddot{K} - 3HK\dot{K}}{K^2} + (3\mathcal{A}' - 16\mathcal{C} - 10\mathcal{D})(\dot{K} + 3HK) + 12\mathcal{C}\dot{H} + 6(6\mathcal{C} - \mathcal{A}')H^2 &\nonumber\\
\phantom{0}- 6\mathcal{B}H\dot{\phi} - \mathcal{B}'\dot{\phi}^2 - 2\mathcal{B}\ddot{\phi} - 2\mathcal{V}' + \frac{3k}{A^2K^2}[2\mathcal{A}'K^2 + (\mathcal{A}' + 2\mathcal{D})(\dot{K} - HK)] &= 0\,,
\end{align}
\end{subequations}
which differ from the previous case only by the additional terms linear in \(k\). For all branches, we find that for the particular class of theories defined by the relation \(\mathcal{A}' = 2\mathcal{C} = -2\mathcal{D}\) the scalar \(K\) decouples from the field equations. This does not come as a surprise, since in this case the theory reduces to the symmetric teleparallel equivalent of scalar-curvature gravity~\cite{Jarv:2018bgs}. Finally, setting \(\mathcal{A}(\phi) = f'(\phi)\) and \(\mathcal{V}(\phi) = (f(\phi) - \phi f'(\phi))/2\), together with \(\mathcal{B} = \mathcal{C} = \mathcal{D} = 0\), one finds that the scalar field equation imposes the condition \(\phi = Q\), while the metric equations reduce to those of $f(Q)$ gravity discussed in section~\ref{ssec:fq} for all branches. Again this result is expected, as the theory is the scalar-nonmetricity representation of $f(Q)$ gravity~\cite{Jarv:2018bgs}. This concludes our discussion of symmetric teleparallel gravity theories and their cosmological dynamics.

\section{Conclusion}\label{sec:conclusion}
We have determined the most general homogeneous and isotropic symmetric teleparallel geometries. We have found a single spatially curved (\(k \neq 0\)) solution to the symmetry conditions, as well as three distinct branches of spatially flat (\(k = 0\)) geometries, each of which arises as a particular limit with \(k \to 0\) from the common spatially curved case. Each of these branches is parametrized by two scalar functions of time, the cosmological scale factor \(A\) in the metric and another scalar \(K\) parametrizing the symmetric teleparallel connection, while a third function, the lapse \(N\), can be absorbed by a choice of the time coordinate. For each branch, we determined the coincident gauge, in which the coefficients of the symmetric teleparallel connection vanish. We found that only for one branch the metric retains the FLRW form in the coincident gauge. Our results show that the common assumption in symmetric teleparallel cosmology, which simultaneously imposes the coincident gauge and the FLRW metric, restricts the geometry to this particular branch, and does not allow for studying the remaining branches. This is due to the fact that both the coincident gauge condition of vanishing connection coefficients and the FLRW form of the metric single out particular classes of coordinate systems, and these two classes may, in general, not have a common member.

Further, we derived the cosmological field equations of a number of classes of symmetric teleparallel gravity theories for all aforementioned branches allowed by the cosmological symmetry. We found several subclasses of theories, including general \(f(Q)\) gravity as well as subclasses of newer general relativity and scalar-nonmetricity gravity, in which the scalar \(K\) in the connection decouples from the metric field equations in the aforementioned coincident / FLRW branch. For these theories, our result reproduces the previously obtained results on symmetric teleparallel cosmology. Moreover, our results also show that this decoupling from the metric field equations constitutes only a particular sector of symmetric teleparallel cosmology, and that \(K\) in general enters also into the metric field equations if a different branch is chosen, or a more general gravity theory is considered, which does not fall into the aforementioned classes. This shows that symmetric teleparallel gravity theories allow for a significantly richer spectrum of cosmological dynamics than has been considered so far in the literature. The physical implications from these modified dynamics depend, of course, on the particular theory under consideration. Due to the plethora of theories discussed in the literature, studying these implications would by far exceed the scope of this article, in which we discuss only general classes of theories. We therefore leave these physical implications for future studies.

We also studied the conservation of energy-momentum-hypermomentum in the aforementioned symmetric teleparallel cosmological backgrounds. The most general matter model compatible with the assumed homogeneity and isotropy is a hyperfluid, and we found that its dynamics, in general, depends on the choice of the particular background geometry and involves the scalar \(K\), unless the hypermomentum vanishes, i.e., in the case of vanishing matter coupling to the symmetric teleparallel connection. This result is independent of the gravity theory under consideration, and this in particular also applies to the class of theories in which \(K\) decouples from the metric field equations, and its evolution remains undetermined by the remaining equations. Further investigations are necessary to determine whether this hints towards a possible pathological dynamics, or a strong coupling issue as it is eminent also in $f(T)$ metric teleparallel gravity theories~\cite{Jimenez:2020ofm,Golovnev:2020nln,Golovnev:2020zpv,Jimenez:2021hai}.

We finally remark that both the gravitational and matter dynamics we presented are independent of any choice of coordinates, as they involve only scalar quantities and their Lie derivatives with respect to the unique future unit vector field obeying the cosmological symmetry, which is canonically defined by the FLRW geometry. In particular, they are independent of the lapse function \(N\), so that the latter may be chosen arbitrarily, irrespective of the theory under consideration; there is no breaking of diffeomorphism invariance which would obstruct this freedom. For the specific branch in which the metric retains its FLRW form also in the coincident gauge, one may also use the coincident gauge condition of vanishing connection coefficients in order to fix the time coordinate, and hence also the lapse function \(N\). Making this peculiar choice, and thus fixing \(N\) through the scalar \(K\) involved in the transformation to the coincident gauge, it appears that diffeomorphism invariance is broken in theories whose dynamics involve \(K\), and thus ascribe dynamics to \(N\) when this choice is made. However, this apparent breaking of diffeomorphism invariance is simply an artifact of the gauge (coordinate) choice, which is in no way singled out from other common coordinate choices.

\begin{acknowledgments}
The author gratefully acknowledges the full financial support by the Estonian Research Council through the Personal Research Funding project PRG356 and by the European Regional Development Fund through the Center of Excellence TK133 ``The Dark Side of the Universe''.
\end{acknowledgments}

\bibliography{nmcosmo,nm}

\end{document}